\documentclass[prl,a4paper,final,superscriptaddress,longbibliography]{revtex4}
\usepackage{graphicx}
\usepackage{color}
\usepackage{bm}
\usepackage{amsmath,amssymb}
\newcommand{\pa}[0]{
\setlength\unitlength{0.005truecm}
\begin{picture}(120,100)(-10,31)
\linethickness{5\unitlength}
\put(0,0){\framebox( 90, 90){}}
\linethickness{25\unitlength}
\put(0,10){\line(1,0){90}}
\put(0,80){\line(1,0){90}}
\end{picture}
}
\newcommand{\pb}[0]{
\setlength\unitlength{0.005truecm}
\begin{picture}(120,100)(-10,31)
\linethickness{5\unitlength}
\put(0,0){\framebox( 90, 90){}}
\linethickness{25\unitlength}
\put(10,0){\line(0,1){90}}
\put(80,0){\line(0,1){90}}
\end{picture}
}
\newcommand{\pc}[0]{
\setlength\unitlength{0.005truecm}
\begin{picture}(120,100)(-10,31)
\linethickness{5\unitlength}
\put(0,0){\framebox( 90, 90){}}
\linethickness{25\unitlength}
\put(0,80){\line(1,0){90}}
\end{picture}
}
\newcommand{\pd}[0]{
\setlength\unitlength{0.005truecm}
\begin{picture}(120,100)(-10,31)
\linethickness{5\unitlength}
\put(0,0){\framebox( 90, 90){}}
\linethickness{25\unitlength}
\put(80,0){\line(0,1){90}}
\end{picture}
}
\newcommand{\pe}[0]{
\setlength\unitlength{0.005truecm}
\begin{picture}(120,100)(-10,31)
\linethickness{5\unitlength}
\put(0,0){\framebox( 90, 90){}}
\linethickness{25\unitlength}
\put(0,10){\line(1,0){90}}
\end{picture}
}
\newcommand{\pf}[0]{
\setlength\unitlength{0.005truecm}
\begin{picture}(120,100)(-10,31)
\linethickness{5\unitlength}
\put(0,0){\framebox( 90, 90){}}
\linethickness{25\unitlength}
\put(10,0){\line(0,1){90}}
\end{picture}
}
\newcommand{\pg}[0]{
\setlength\unitlength{0.005truecm}
\begin{picture}(120,100)(-10,31)
\linethickness{5\unitlength}
\put(0,0){\framebox( 90, 90){}}
\linethickness{25\unitlength}
\end{picture}
}

\begin{document}

\title{Quantum dimer model containing Rokhsar--Kivelson point expressed by spin-1/2 Heisenberg antiferromagnets}


\author{Yuhei Hirose$^{\ast}$\footnote[0]{$^{\ast}$Email:\;hiro3332@gmail.com}, Akihide Oguchi, and Yoshiyuki Fukumoto} 
\affiliation{Department of Physics, Faculty of Science and Technology, Tokyo University of Science, Noda, Chiba 278-8510, Japan}


\date{\today}

\begin{abstract}
We obtain a quantum dimer model (QDM) containing a Rokhsar--Kivelson (RK) point expressed by spin-1/2 Heisenberg antiferromagnets on a diamond-like decorated square lattice. 
This lattice has macroscopically degenerated nonmagnetic ground states, which are equivalent to the Hilbert space of a square-lattice QDM. 
Then, a square-lattice QDM containing the RK point as a second-order effective Hamiltonian is obtained by introducing further neighbor couplings as perturbation.
Our model can provide a new method for the experimental realization of a resonating valence bond state in the QDM.
\end{abstract}


\maketitle


\section{I.~Introduction}

The quantum spin liquid (QSL) state is an exotic state that has been extensively studied. Owing to a strong quantum effect and frustration, a QSL does not order at absolute zero temperature and does not invoke any spontaneous symmetry breaking\cite{ref.QSL2010,ref.book,ref.QSL2017}. The QSL state originates from the resonating valence bond (RVB) state of two-dimensional spin-1/2 Heisenberg antiferromagnets, as proposed by P.~W.~Anderson in 1973\cite{ref.Anderson1973}. 

Over the past decade, the Kitaev model has been pivotal in the study of QSL states\cite{ref.Kitaev1}.  
This model is an exactly solvable spin-1/2 model on a honeycomb lattice and forms a topological QSL in the ground state, which is described by the Majorana fermion split from the electron spin\cite{ref.Kitaev2}. 
However, its experimental realization has been considered difficult\cite{ref.Kitaev3,ref.Kitaev4,ref.Kitaev4-2,ref.Kitaev5,ref.Kitaev6}, because the Kitaev model contains extremely anisotropic ferromagnetic interactions.
Very recently, a QSL has finally been realized, which has a significant effect on both theoretical and experimental physics. 
Y.~Kasahara et al. reported long-range magnetic order suppression and a field-induced QSL by the application of a magnetic field to $\rm{\alpha}$-$\rm{RuCl_3}$ compounds with a dominant Kitaev interaction, i.e., a bond-dependent Ising-type interaction\cite{ref.Kitaev7}. 

On the other hand, a quantum simulator can also be expected, employing an optical lattice\cite{ref.Bloch2008,ref.Bloch2017,ref.Bhuvanesh,ref.Fukuhara} to realize a model in which the QSL state can be obtained exactly, i.e., in which the wave functions of the QSL state can be given exactly. 
Actually, a one-dimensional Heisenberg model has been realized in a quantum simulator using an optical lattice\cite{ref.Fukuhara}. 
Therefore, it is expected that a simple model containing only isotropic Heisenberg interactions can be realized easily by using such a quantum simulator, as opposed to a model containing a strong anisotropic interaction such as the Kitaev model.
Then, the question of whether such a Heisenberg model exists, that is, one that can provide the QSL state exactly, can be answered. 

In 1988, D.~S.~Rokhsar and S.~A.~Kivelson proposed a quantum dimer model (QDM)\cite{ref.Rokhsar1988} as a phenomenological Hamiltonian of the RVB theory. 
The QDM is given by 
\begin{align}
H_{\rm{QDM}}=&-t\sum \Bigl( \Bigr| \pa{} \Bigl\rangle \Bigl\langle\pb{}\Bigr|+\Bigr| \pb{} \Bigl\rangle \Bigl\langle\pa{}\Bigr| \Bigr)
		      +v\sum \Bigl( \Bigr| \pa{} \Bigl\rangle \Bigl\langle\pa{}\Bigr|+\Bigr| \pb{} \Bigl\rangle \Bigl\langle\pb{}\Bigr| \Bigr),
\label{eq:1}
\end{align}
where $t$ and $v$ represent the pair-hopping amplitude and dimer--dimer interaction, respectively. 
At $v=t>0$, known as the Rokhsar--Kivelson (RK) point, the ground state is exactly an equal-amplitude superposition of the dimer-covering configurations, which can be ascribed to an RVB state\cite{ref.Rokhsar1988}. The numerous studies on the QDM over the years\cite{ref.book,ref.square_qdm_1996,ref.square_qdm_2006,ref.square_qdm_2008,ref.square_qdm_2014,ref.square_qdm_2018,ref.square_qdm_2019,ref.triangular_qdm_2001,ref.triangular_qdm_2005,ref.cubic_qdm_2003a,ref.cubic_qdm_2003b,ref.hexagonal_qdm_2001,ref.hexagonal_qdm_2015} suggest the possibility of finding a Heisenberg model from which a QDM can be derived. 
However, it still has not been clarified whether a QDM can be established from realistic quantum spin systems that only have isotropic Heisenberg interactions. 

\begin{figure}[t] 
\begin{center}
\includegraphics[width=10.0cm,clip]{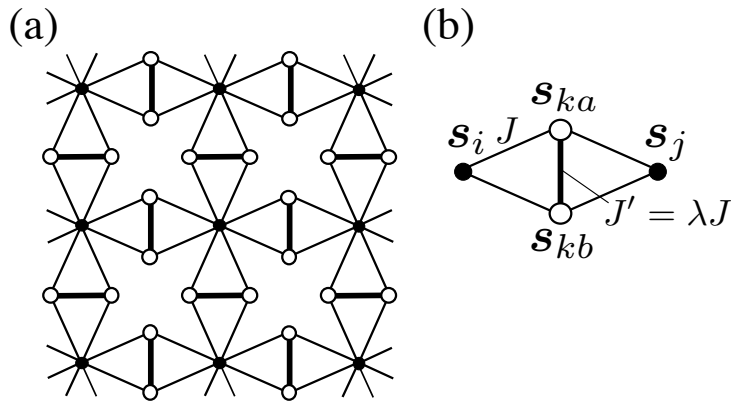}
\end{center}
\caption{ Structure of (a) the DDSL and (b) the diamond unit. The thin and thick solid lines represent the antiferromagnetic interactions $J$ and $J'=\lambda J$, respectively. We denote $\bm{s}_i$ and $\bm{s}_j$ as the edge spins and the pair ($\bm{s}_{ka}, \bm{s}_{kb}$) as a bond spin pair. The edge spins and bond spin pairs are indicated by the closed and open circles, respectively. The magnitude of all spin operators is $1/2$.  
 }
\label{fig:1}
\end{figure}

In this paper, we report the possibility of obtaining a QDM containing an RK point from isotropic spin-1/2 Heisenberg antiferromagnets on a diamond-like decorated square lattice (DDSL) perturbed by further neighbor couplings.  
In the DDSL the bonds of a square lattice are replaced by diamond units\cite{ref.Strecka}, as shown in Fig.~\ref{fig:1}(a). 

If the interaction strength of the four sides of a diamond unit is denoted as $J$ and the diagonal bond is denoted as $J'=\lambda J$, 
then their ratio $\lambda$ determines the ground-state properties\cite{ref.Hirose1}.  
As shown in Fig.~\ref{fig:1}(b), we denote the four spin-1/2 operators in a diamond unit as $\bm{s}_i$, $\bm{s}_j$, $\bm{s}_{ka}$, and $\bm{s}_{kb}$. 
Thus, the Hamiltonian can be written as 
\begin{align} 
H=&\sum_{\langle i,j \rangle}h_{i,j}
 \label{eq:2},\\
h_{i,j}=&J(\bm{s}_i+\bm{s}_j)\cdot(\bm{s}_{ka}+\bm{s}_{kb})
+J^{\prime}\left(\bm{s}_{ka}\cdot\bm{s}_{kb}+\frac{3}{4}\right)
  \label{eq:3},
\end{align}
where $\langle i,j \rangle$ represents a nearest-neighbor pair of the square lattice. 
Here, $\bm{s}_i$ and $\bm{s}_j$ are known as the edge spins (closed circles in Fig.~\ref{fig:1}(b)) and the pair ($\bm{s}_{ka}, \bm{s}_{kb}$) is known as a bond spin pair (open circles). 
It should be noted that in Eq.~(\ref{eq:3}) the energy of the bond spin pair is measured from that of the singlet dimer. 

\begin{figure}[t] 
\begin{center}
\includegraphics[width=10.0cm,clip]{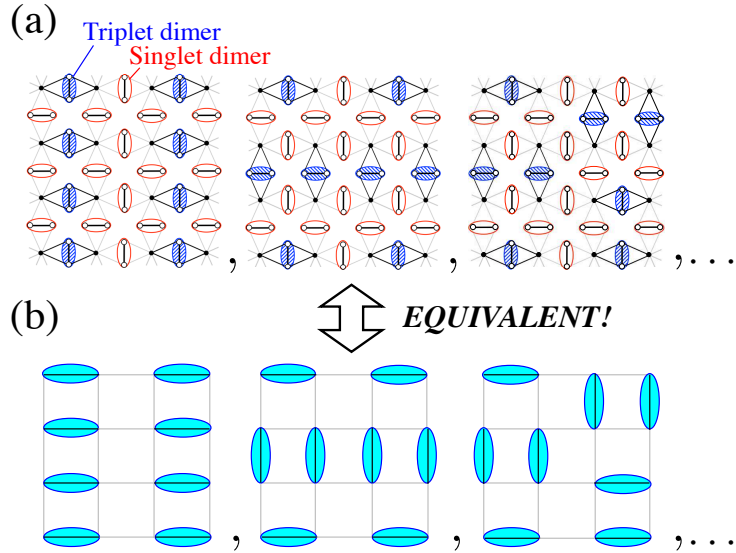}
\end{center}
\caption{(Color online) (a) Ground states on the DDSL with $0.974<\lambda<2.0$, exhibiting a nontrivial macroscopic degeneracy. The shaded blue and empty red ovals represent the triplet and singlet dimers on the bond spin pair, respectively. (b) Square-lattice dimer-covering states. There is no interaction between dimers and each dimer is independent. 
It can be seen that the arrangements on panels (a) and (b) are equivalent.   
 }
\label{fig:2}
\end{figure}

For the ground states of the DDSL with $0.974<\lambda<2.0$, the present system exhibits a nontrivial macroscopic degeneracy\cite{ref.Hirose1,ref.Morita}. 
As shown in Fig.~\ref{fig:2}(a), in these ground states the diamonds with triplet dimers (shaded blue ovals) and with singlet dimers (empty red ovals) are arranged such that the number of diamonds with triplet dimers is the largest, and two or more diamonds with triplet dimers are not located next to each other, according to the variational method and the Lieb--Mattis theorem\cite{ref.Hirose1}.  
Then, as shown in Fig.~\ref{fig:2}(b), we demonstrate that the ground states in Fig.~\ref{fig:2}(a) are equivalent to the square-lattice dimer-covering states. 
First, by considering the diamond with a singlet dimer in Fig.~\ref{fig:2}(a), we obtain 
\begin{align}
J(\bm{s}_i+\bm{s}_j)\cdot(\bm{s}_{ka}+\bm{s}_{kb})|\sigma,\sigma^{\prime}\rangle_{i,j}|s\rangle_{k}=0
\label{eq:4-0},
\end{align}
where $\sigma,\sigma^{\prime}=\uparrow$ or $\downarrow$ and $|s\rangle_{k}=\left(|\!\uparrow_{ka},\downarrow_{kb}\rangle-|\!\downarrow_{ka},\uparrow_{kb}\rangle\right)/\sqrt{2}$ represents the singlet dimer of a bond spin pair. 
Equation~(\ref{eq:4-0}) indicates that due to the singlet dimer on the bond spin pairs the Hamiltonian in Eq.~(\ref{eq:3}) effectively vanish. 
Therefore, we can regard the diamond with a singlet dimer as no dimer in the QDM.  
On the other hand, the eigenvector of the diamond with triplet dimer can be written as
 \begin{align}
   |\phi^g\rangle_{i,j,k}=\frac{1}{\sqrt{3}}
   \left(
   |T^+\rangle_{i,j}|t^-\rangle_k+|T^-\rangle_{i,j}|t^+\rangle_k
   -|T^0\rangle_{i,j}|t^0\rangle_k
   \right),
\label{eq:4} 
\end{align}
where $\{|T^+\rangle_{i,j}, |T^0\rangle_{i,j}, |T^-\rangle_{i,j}\}$ and $\{|t^+\rangle_{k}, |t^0\rangle_{k}, |t^-\rangle_{k}\}$ represent the triplet dimers of edge spins and of a bond spin pair, respectively, i.e., 
\begin{equation}
 |T^{\alpha}\rangle_{i,j}=\begin{cases}
|\!\uparrow_{i},\uparrow_{j}\rangle & \mbox{ $(\alpha=+)$} \\
\left(|\!\uparrow_{i},\downarrow_{j}\rangle+|\!\downarrow_{i},\uparrow_{j}\rangle\right)/\sqrt{2}   & \mbox{ $(\alpha=0)$} \\
|\!\downarrow_{i},\downarrow_{j}\rangle           &  \mbox{ $(\alpha=-)$}
\end{cases}
\label{eq:4-2}
\end{equation}
and 
\begin{equation}
 |t^{\alpha}\rangle_{k}=\begin{cases}
|\!\uparrow_{ka},\uparrow_{kb}\rangle & \mbox{ $(\alpha=+)$} \\
\left(|\!\uparrow_{ka},\downarrow_{kb}\rangle+|\!\downarrow_{ka},\uparrow_{kb}\rangle\right)/\sqrt{2}   & \mbox{ $(\alpha=0)$} \\
|\!\downarrow_{ka},\downarrow_{kb}\rangle         &  \mbox{ $(\alpha=-)$}
\end{cases}
\label{eq:4-3},
\end{equation}
and the eigenvalue in Eq.~(\ref{eq:4}) is $J(\lambda-2)$.  
Equation~(\ref{eq:4}) can also be written as 
 \begin{align}
   |\phi^g\rangle_{i,j,k}=\frac{1}{\sqrt{3}}
   \left(
   |s\rangle_{ka,i}|s\rangle_{kb,j}+|s\rangle_{ka,j}|s\rangle_{kb,i}
   \right),
\label{eq:4-4} 
\end{align}
where $|s\rangle_{k\alpha,l}=\left(|\!\uparrow_{k\alpha},\downarrow_{l}\rangle-|\!\downarrow_{k\alpha},\uparrow_{l}\rangle\right)/\sqrt{2}$ ($\alpha=a,b$ and $l=i,j$) and Eq.~(\ref{eq:4-4}) is a plaquette RVB state. 
As
\begin{align}
(\bm{s}_i+\bm{s}_j+\bm{s}_{ka}+\bm{s}_{kb})^2|\phi^g\rangle_{i,j,k}=0
\label{eq:4-5}
\end{align}
is obtained, we find that Eqs.~(\ref{eq:4}) and (\ref{eq:4-4}) describe nonmagnetic tetramer-singlet states. 
Therefore, the ground states shown in Fig.~\ref{fig:2}(a), known as macroscopically degenerated tetramer-dimer (MDTD) states, are nonmagnetic, because they consist only of tetramer-singlet states and singlet dimers.  
\begin{figure}[t] 
\begin{center}
\includegraphics[width=10.0cm,clip]{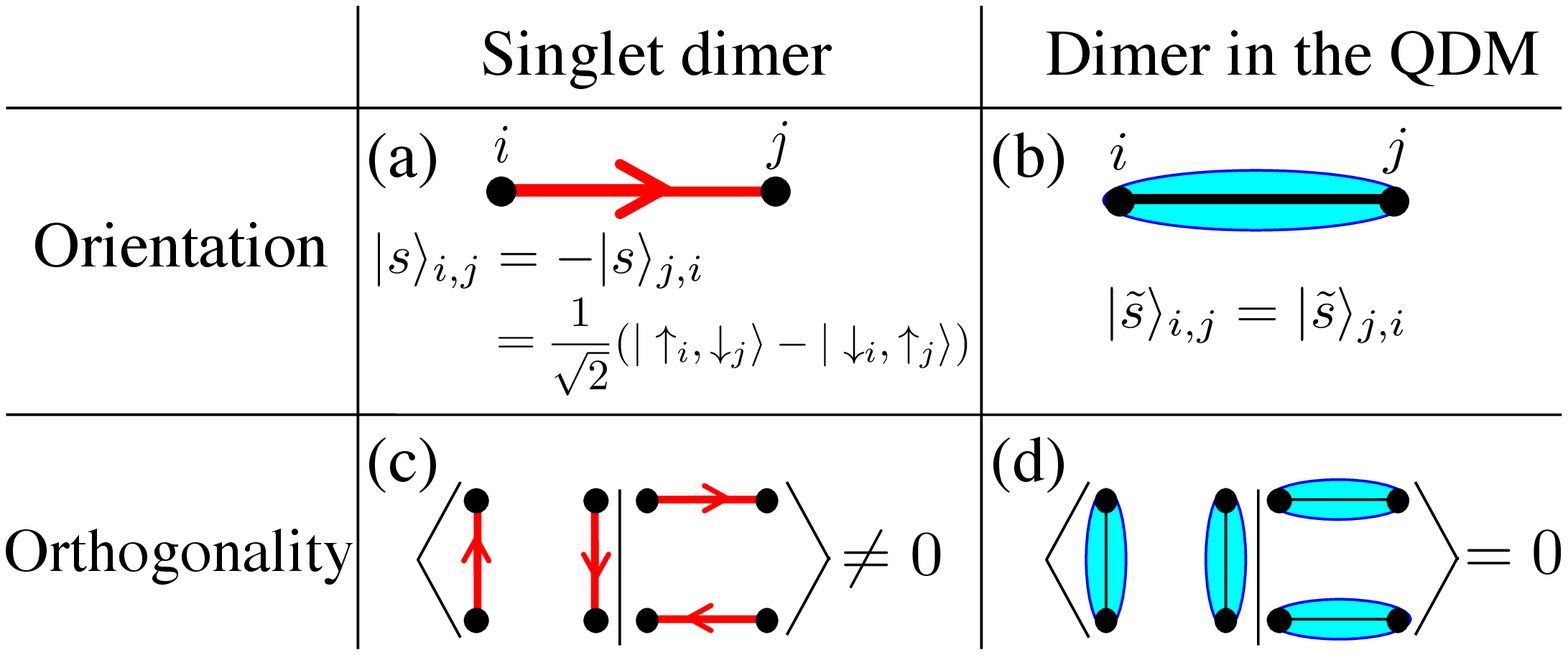}
\end{center}
\caption{(Color online) Properties of the singlet dimer and dimer in the QDM.  
 }
\label{fig:3}
\end{figure}
\begin{figure}[h] 
\begin{center}
\includegraphics[width=15.0cm,clip]{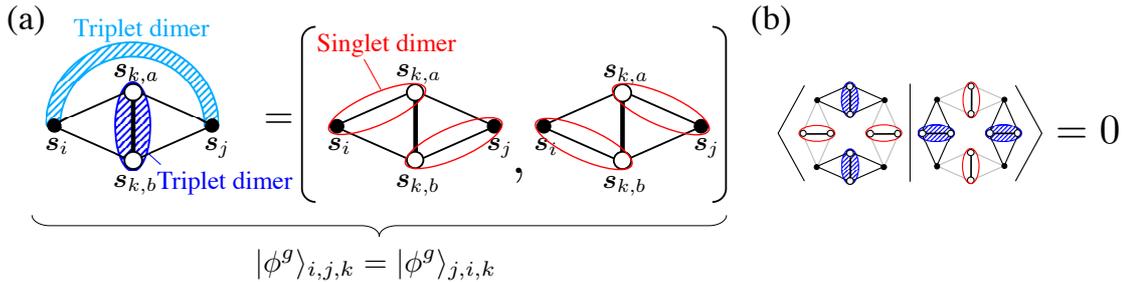}
\end{center}
\caption{(Color online) Properties of the tetramer-singlet states: (a) the tetramer-singlet, equivalent to the plaquette RVB state, has no orientation and (b) the different tetramer-singlet covering states are orthogonal. 
 }
\label{fig:4}
\end{figure}

Here, it should be noted that the properties of the singlet dimer and the dimer of QDM are very different. 
The first difference is that the singlet dimer $|s\rangle_{i,j}$ has an orientation (choice of a sign), i.e., $|s\rangle_{i,j}=-|s\rangle_{j,i}=(|\uparrow_i,\downarrow_j\rangle-|\downarrow_i,\uparrow_j\rangle)/\sqrt{2}$. 
Contrary to this, the dimer in the QDM has no orientation and it can be assumed that $|\tilde{s}\rangle_{i,j}=|\tilde{s}\rangle_{j,i}$ (see Fig.~\ref{fig:3}(a) and~(b)). 
The second difference is that each of the singlet dimer covering states is non-orthogonal as shown in Fig.~\ref{fig:3}(c). The orthonormal basis $|m\rangle$ (Fig.~\ref{fig:3}(d)) is obtained by using the mathematical formula $|\Psi_m\rangle=\sum_{n}(S^{-1/2})_{mn}|\Psi_n\rangle$, where $|\Psi_n\rangle$ are the dimer coverings and $S_{mn}=\langle\Psi_m|\Psi_n\rangle$. However, the calculation of $(S^{-1/2})_{mn}$ is very difficult\cite{ref.Raman,ref.Vernay}.

\begin{figure}[h] 
\begin{center}
\includegraphics[width=8.0cm,clip]{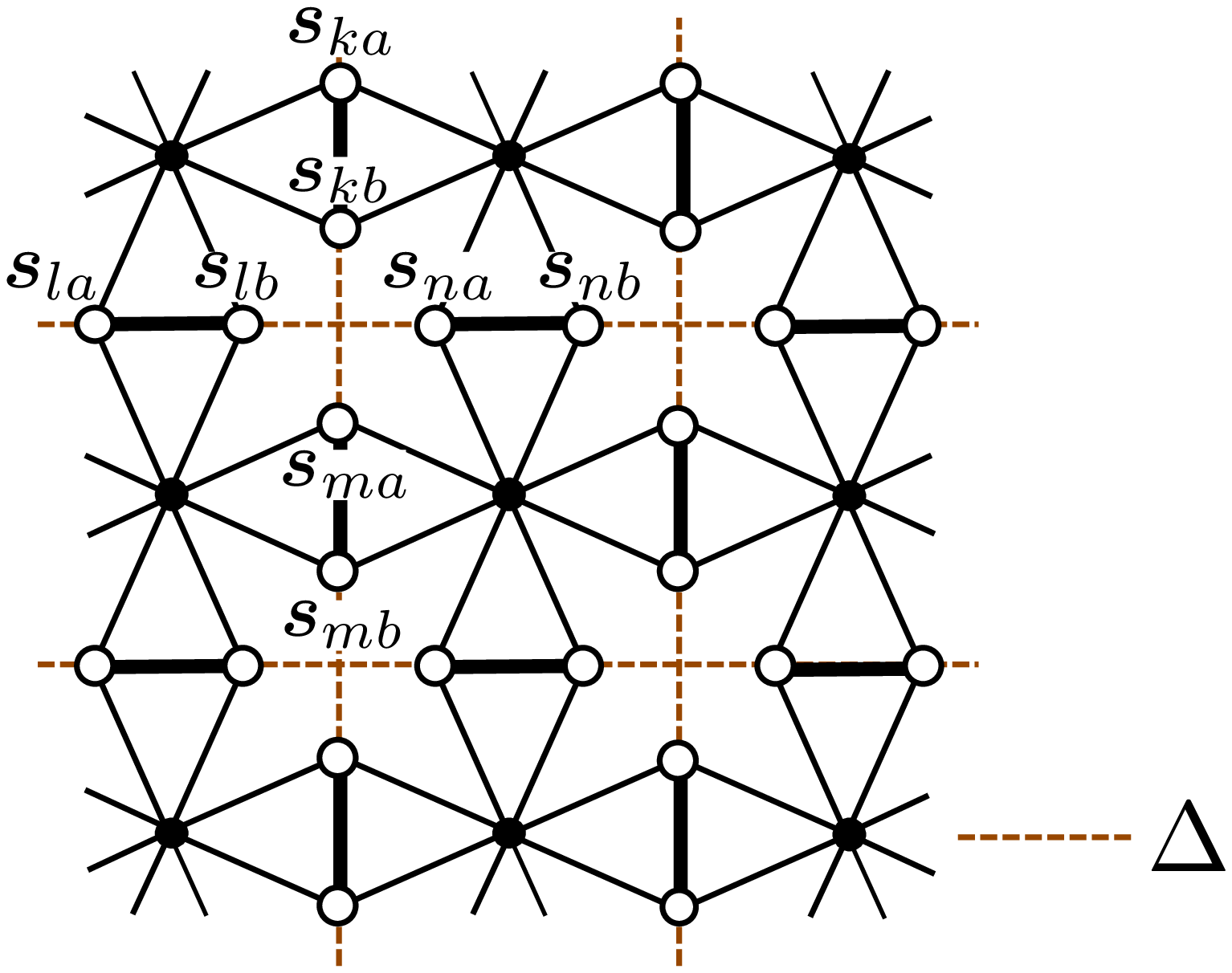}
\end{center}
\caption{(Color online) Structure of the DDSL with the introduction of further neighbor couplings $\Delta$. }
\label{fig:5}
\end{figure}
The tetramer-singlet state in this study has no orientation, i.e., $|\phi^g\rangle_{i,j,k}=|\phi^g\rangle_{j,i,k}$ holds (Fig.~\ref{fig:4}(a)), and each of the tetramer-singlet covering states are orthogonal (Fig.~\ref{fig:4}(b)). Furthermore, each tetramer-singlet state is independent. 
These properties enable us to obtain an isotropic Heisenberg Hamiltonian, in which the ground-state manifold coincides with the classical close-packed dimer-covering states on any lattices, if the Hamiltonian in Eq.~(3) is applied on every link of the lattice. 
Therefore, as shown in Fig.~\ref{fig:2}(b), the MDTD states on the DDSL are exactly the same as the square-lattice dimer-covering states.

However, as the magnitude of the bond spin pair is a conserved quantity, i.e., $[(\bm{s}_{ka}+\bm{s}_{kb})^2,h_{i,j}]=0$, there is no hopping of the tetramer-singlet state; thus, there is no transition between each state in the MDTD states. 
Therefore, in this study, we show that the hopping of the tetramer-singlet states can be achieved by introducing further neighbor couplings, $\Delta$, as perturbation (Fig.~\ref{fig:5}), and the obtained square-lattice QDM as a second-order effective Hamiltonian contains the RK point. 
It should be noted that we previously obtained a square-lattice QDM by an alternative method of introducing further neighbor couplings\cite{ref.Hirose2,ref.Hirose3}. However, we obtained $v\le0$ in the entire region of $\lambda$, which indicated that the obtained QDM did not contain the RK point. Furthermore, we also obtained a QDM by introducing another kind of further neighbor coupling in addition to $\Delta$ as shown in Fig.~5 of the present paper (see $\Delta_{\rm{I}}$ in Fig.~3 in Ref.~\onlinecite{ref.Hirose4}). However, we found that the introduction of $\Delta_{\rm{I}}$ did not provide the RK point because $\Delta_{\rm{I}}$ resulted in a significantly larger pair-hopping amplitude than the dimer--dimer repulsive interaction, i.e., $t(\gg v)$\cite{ref.Hirose4}. 

This paper is organized as follows.
In Sect.~II., we define the perturbation Hamiltonian and obtain a square-lattice QDM as a second-order effective Hamiltonian. 
In Sect.~III., we describe the details of the second-order perturbation processes. 
In Sect.~IV., we discuss the calculation results described in Sect.~3 and demonstrate that our effective Hamiltonian as a square-lattice QDM contains the RK point. 
In Sect.~V., we summarize the obtained results. 

\section{II.~Perturbation Hamiltonian and square-lattice QDM as a second-order effective Hamiltonian}

The perturbation Hamiltonian can be written as
\begin{align}
H^{\prime}=\sum_{\langle k,m\rangle}V_{k,m}+\sum_{\langle l,n\rangle}V_{l,n}
\label{eq:5},
\end{align}
\begin{align}
V_{k,m}=\Delta_{k,m}\bm{s}_{kb}\cdot\bm{s}_{ma},\;V_{l,n}=\Delta_{l,n}\bm{s}_{lb}\cdot\bm{s}_{na}
\label{eq:6},
\end{align}
where $\Delta_{k,m}=\Delta_{l,n}=\Delta$ while $\langle k,m\rangle$ and $\langle l,n\rangle$ represent the pair facing each other on one plaquette, as shown in Fig.~\ref{fig:5}. 

Then, based on Eq.~(\ref{eq:5}), we adopt perturbation energies up to the second order. 
First, we find that the first-order perturbation energy is zero, which is described in Sect.~III. 
Therefore, we only consider the second-order perturbation and obtain a second-order effective Hamiltonian. 
It should be noted that an arbitrary second-order process cannot be created by perturbation bonds belonging to different plaquettes. 
As the possible second-order processes are created using the perturbation bonds on a plaquette twice, the second-order effective Hamiltonian can be written by
\begin{align}
   H_{\rm{eff}}=&-t^{(=)}\sum\Bigr| \pa{}\Bigl\rangle \Bigl\langle\pb{}\Bigr|-t^{(||)}\sum\Bigr| \pb{} \Bigl\rangle \Bigl\langle\pa{}\Bigr|+\epsilon_2^{(=)}\sum \Bigr| \pa{} \Bigl\rangle \Bigl\langle\pa{}\Bigr|+\epsilon_2^{(||)}\sum\Bigr| \pb{} \Bigl\rangle \Bigl\langle\pb{}\Bigr| \notag \\
   &+\epsilon_1^{(\rm{up})}\sum \Bigr| \pc{} \Bigl\rangle \Bigl\langle\pc{}\Bigr|+\epsilon_1^{(\rm{down})}\sum\Bigr| \pe{} \Bigl\rangle \Bigl\langle\pe{}+\epsilon_1^{(\rm{right})}\sum \Bigr| \pd{} \Bigl\rangle \Bigl\langle\pd{}\Bigr|+\epsilon_1^{(\rm{left})}\sum\Bigr| \pf{} \Bigl\rangle \Bigl\langle\pf{}\Bigr|\notag\\
   &+\epsilon_0\sum\Bigr| \pg{} \Bigl\rangle \Bigl\langle\pg{}\Bigr|
\label{eq:7}, 
\end{align}
where $t^{(=)}$ and $t^{(||)}$ represent the second-order pair-hopping amplitudes of dimers and $\epsilon_2^{(=)}$, $\epsilon_2^{(||)}$, $\epsilon_1^{(\rm{up})}$, $\epsilon_1^{(\rm{down})}$, $\epsilon_1^{(\rm{right})}$, $\epsilon_1^{(\rm{left})}$, and $\epsilon_0$ represent the second-order perturbation energies, respectively. Henceforth, the dimer refers to the tetramer-singlet state $\phi^g$.
The summations are taken over the plaquettes of the lattice.  
Furthermore, considering the translational and rotational symmetries of the system, it can be written that 
\begin{align}
t=&t^{(=)}=t^{(||)}\notag \\
\epsilon_2=&\epsilon_2^{(=)}=\epsilon_2^{(||)}\notag \\
\epsilon_1=&\epsilon_1^{(\rm{up})}=\epsilon_1^{(\rm{down})}=\epsilon_1^{(\rm{right})}=\epsilon_1^{(\rm{left})}
\label{eq:7-2}. 
\end{align}
Therefore, Eq.~(\ref{eq:7}) can be obtained by
\begin{equation}
H_{\rm{eff}}=-t \hat{T}+\epsilon_2\hat{D}_2+\epsilon_1\hat{D}_1+\epsilon_0\hat{D}_0
\label{eq:7-3}, 
\end{equation}
where
\begin{align}
\hat{T}=&\sum \Bigl( \Bigr| \pa{}\Bigl\rangle \Bigl\langle\pb{}\Bigr|+\Bigr| \pb{} \Bigl\rangle \Bigl\langle\pa{}\Bigr| \Bigr),\notag \\
\hat{D}_2=&\sum \Bigl( \Bigr| \pa{} \Bigl\rangle \Bigl\langle\pa{}\Bigr|+\Bigr| \pb{} \Bigl\rangle \Bigl\langle\pb{}\Bigr| \Bigr),\notag \\
\hat{D}_1=&\sum \Bigl( \Bigr| \pc{} \Bigl\rangle \Bigl\langle\pc{}\Bigr|+\Bigr| \pe{} \Bigl\rangle \Bigl\langle\pe{}\Bigr|+\Bigr| \pd{} \Bigl\rangle \Bigl\langle\pd{}\Bigr|+\Bigr| \pf{} \Bigl\rangle \Bigl\langle\pf{}\Bigr|\Bigr),\notag\\
\hat{D}_0=&\sum  \Bigr| \pg{} \Bigl\rangle \Bigl\langle\pg{}\Bigr|.
\label{eq:8}
\end{align}
As $H_{\rm{eff}}$ conserves the number of dimers, we can write
\begin{align}
\hat{D}_2+\hat{D}_1+\hat{D}_0=N=\mbox{[total number of plaquettes]}
\label{eq:8-2}
\end{align}
and 
\begin{align}
\frac{1}{2}\left(2\hat{D}_2+\hat{D}_1\right)=\frac{N}{2}=\mbox{[total number of dimers]}
\label{eq:8-3}, 
\end{align}
where the coefficient 1/2 on the left-hand side of Eq.~(\ref{eq:8-3}) is introduced to prevent the double counting of dimers. 
Using Eqs.~(\ref{eq:8-2}) and (\ref{eq:8-3}) and eliminating $\hat{D}_1$ and $\hat{D}_0$ from $H_{\rm{eff}}$, Eq.~(\ref{eq:7-3}) can be rewritten as 
\begin{equation}
H_{\rm{eff}}= -t \hat{T}+(\epsilon_2-2\epsilon_1+\epsilon_0)\hat{D}_2+\epsilon_1N
\label{eq:9}. 
\end{equation}
Here, the coefficient of $\hat{D}_2$ on the right-hand side of Eq.~(\ref{eq:9}) represents the dimer--dimer interaction
\begin{equation}
v=\epsilon_2-2\epsilon_1+\epsilon_0
\label{eq:10}  
\end{equation}
and the sum of the first and second terms on the right-hand side of Eq.~(\ref{eq:9}) is $H_{\rm{QDM}}$. 
It should be noted that Eq.~(\ref{eq:10}) becomes repulsive (attractive) when there is a large (small) energy gain of a plaquette with one dimer, i.e., $|\epsilon_1|\gg|\epsilon_0|, |\epsilon_2|$ ($|\epsilon_1|\ll|\epsilon_0|, |\epsilon_2| $) because $\epsilon_2$, $\epsilon_1$, and $\epsilon_0$ are all negative.   

\section{III.~Second-order perturbation processes}

\begin{center}
\begin{figure*}[t]
\includegraphics[width=18cm,clip]{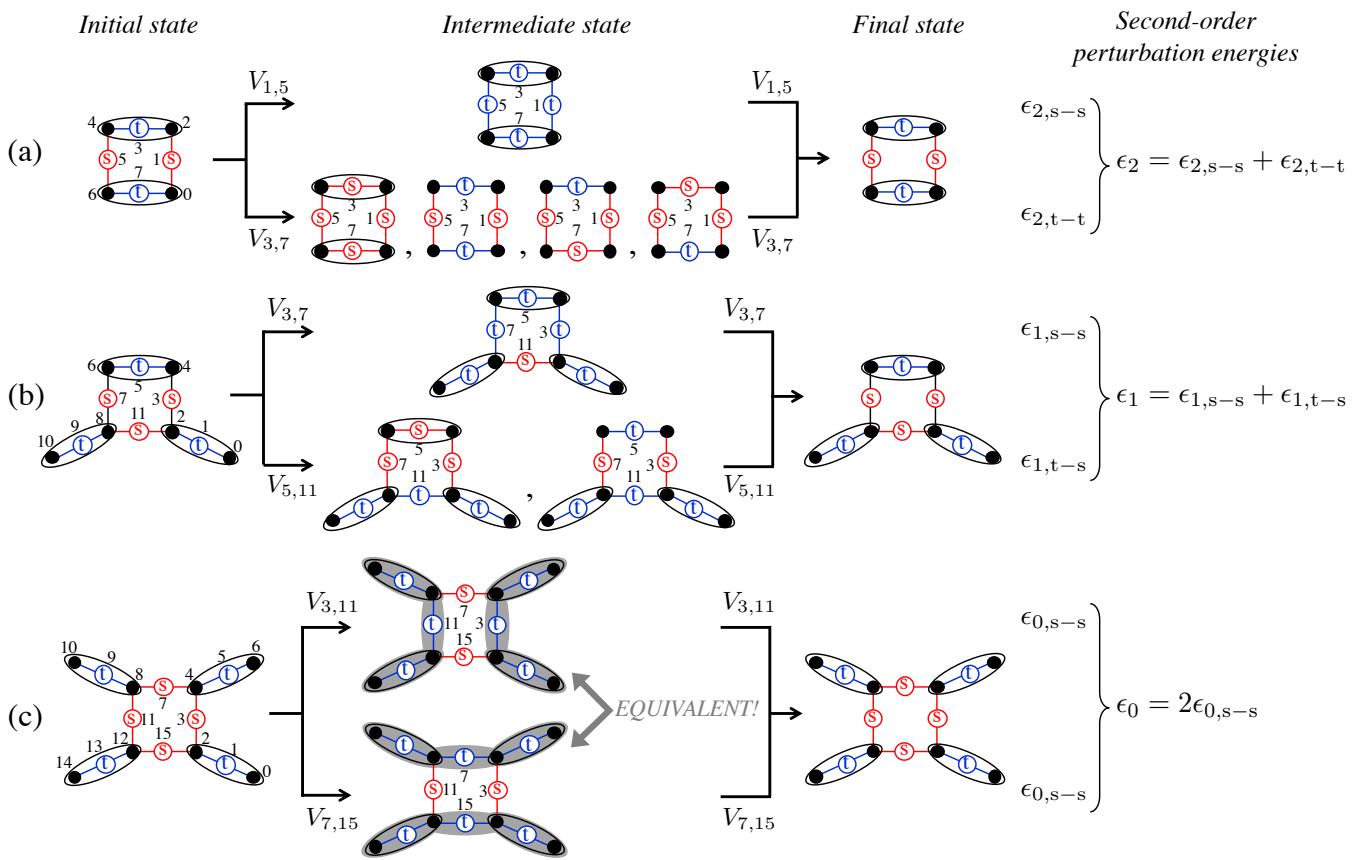}
\caption{(Color online) Second-order perturbation processes and energies generating the dimer--dimer interaction $v$. The initial state has (a) two dimers, (b) one dimer, and (c) no dimer on the plaquette. The blue (red) bonds indicate that the bond spin pair is in triplet (singlet) states. A blue (red) bonds in the ovals represent the tetramer-singlet state $\phi^g$ (the state that is obtained by replacing $|t\rangle$ in $\phi^g$ by $|s\rangle$).}
\label{fig:6}
\end{figure*}
\end{center}

The possible second-order perturbation processes that generate the dimer--dimer interaction $v$ are shown in Fig.~\ref{fig:6}.
As shown in Fig.~\ref{fig:6}(a), the initial state has two dimers on the plaquette. 
There are two kinds of processes, which are produced by $V_{1,5}$ and $V_{3,7}$. 
When operator $V_{1,5}$ is applied, the singlet states at sites $1$ and $5$ turn into triplet states in the intermediate state. 
On the other hand, when operator $V_{3,7}$ is applied, four kinds of clusters are generated in the intermediate state. 
Defining the second-order perturbation energy as $\epsilon_{2,\rm{s-s}}$ ($\epsilon_{2,\rm{t-t}}$) when operator $V_{1,5}$ ($V_{3,7}$) is applied we can write 
\begin{align}
\epsilon_2=\epsilon_{2,\rm{s-s}}+\epsilon_{2,\rm{t-t}}
\label{eq:11}.
\end{align} 
It should be noted that when a perturbation bond operates on a dimer-covering state, clusters with the edge spins that become ``defects" are generated in the intermediate state, 
where a defect refers to an edge spin that belongs to two or more dimers or to no dimer. No perturbation bond on another plaquette can remove these defects, which holds for all second-order perturbation processes described in the following.
As shown in Fig.~\ref{fig:6}(b), the initial state has one dimer on the plaquette. 
When operator $V_{3,7}$ ($V_{5,11}$) is applied, one cluster (two kinds of clusters) is (are) generated in the intermediate state. 
Defining the second-order perturbation energy as $\epsilon_{1,\rm{s-s}}$ ($\epsilon_{1,\rm{t-s}}$) when operator $V_{3,7}$ ($V_{5,11}$) is applied, we can write 
\begin{align}
\epsilon_1=&\epsilon_{1,\rm{s-s}}+\epsilon_{1,\rm{t-s}}
\label{eq:12}. 
\end{align}
As shown in Fig.~\ref{fig:6}(c), the initial state has no dimer on the plaquette. In this case, the clusters formed in the intermediate state become equivalent. 
Therefore, we have the same contributions from the process when $V_{3,11}$ and $V_{7,15}$ are applied. 
Thus, defining the second-order perturbation energy as $\epsilon_{0,\rm{s-s}}$, when operator $V_{3,11}$ or $V_{7,15}$ is applied we can write 
\begin{align}
\epsilon_0=&2\epsilon_{0,\rm{s-s}}
\label{eq:13}. 
\end{align}

Here, we discuss the first-order perturbation energy. It can be seen that the upper process shown in Fig.~\ref{fig:6}(a) and the upper and lower processes shown in Figs.~\ref{fig:6}(b) and (c) do not give a first-order term of the effective Hamiltonian because these intermediate states do not contain diagonal terms. 
On the other hand, the intermediate state of the lower process shown in Fig.~\ref{fig:6}(a) appears to contain diagonal terms corresponding to the second state from the left. 
However, similarly to this case, it can be shown easily that the first-order term is zero, as follows. When we consider the tetramer-singlet state $|\phi^g\rangle_{i,j,k}$, the expectation of the spin operator $s_{k,\nu}^{\xi}$ $(\xi=x,y,z$ ; $\nu=a,b)$ of the state $|\phi^g\rangle_{i,j,k}$ can be obtained as $\langle\phi^g_{i,j,k}|s_{k,\nu}^{\xi}|\phi^g_{i,j,k}\rangle=0$. Therefore, we obtain $\langle\phi^g_{i,j,k};\phi^g_{i',j',k'}|V_{k,k'}|\phi^g_{i,j,k};\phi^g_{i',j',k'}\rangle=0$ and find that the first-order term of the effective Hamiltonian is not provided.

Substituting Eqs.~(\ref{eq:11})--(\ref{eq:13}) into Eq.~(\ref{eq:10}), we finally obtain the dimer--dimer interaction, $v$. 
\begin{center}
\begin{figure*}[h]
\includegraphics[width=17cm,clip]{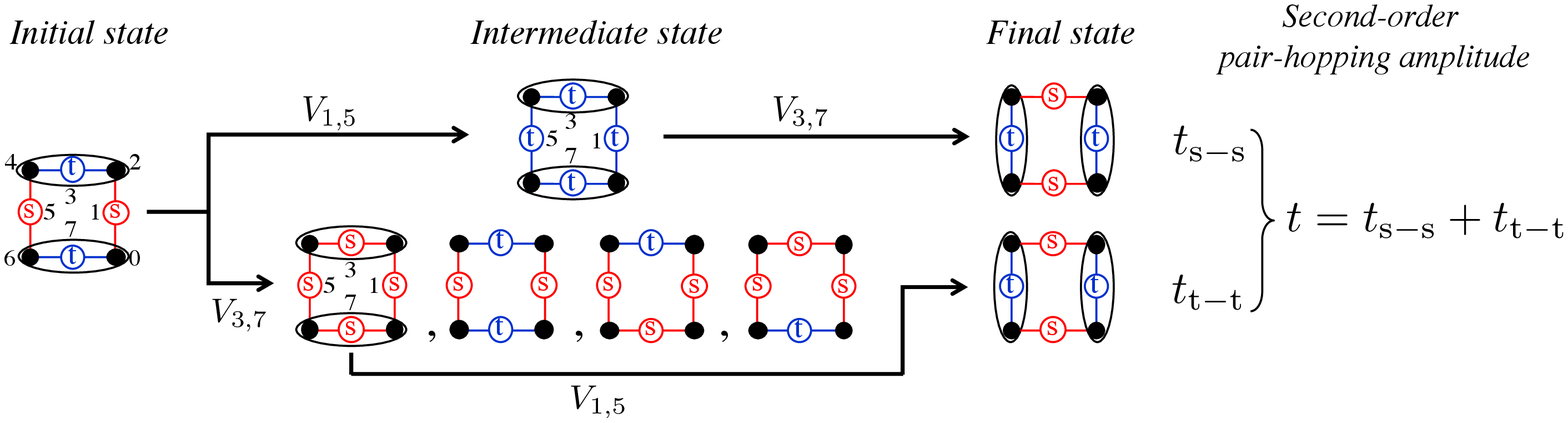}
\caption{(Color online) Second-order perturbation processes for the pair-hopping of dimers. There are two kinds of processes.}
\label{fig:7}
\end{figure*}
\end{center}

Next, we consider possible second-order processes for the pair-hopping of dimers. 
As shown in Fig.~\ref{fig:7}, there are two kinds of processes. 
In the case of the upper process shown in Fig.~\ref{fig:7}, a state where all bond spin pairs at sites $1$, $3$, $5$, and $7$ are triplet states is generated in the intermediate state. 
Furthermore, when $V_{3,7}$ operates on the intermediate state, and the triplet states at sites $3$ and $7$ become singlet states, the pair-hopping of dimers occurs. 
On the other hand, in the case of the lower process shown in Fig.~\ref{fig:7}, one of the intermediate states is the state where all bond spin pairs at sites $1$, $3$, $5$, and $7$ are singlet states. Furthermore, when $V_{1,5}$ operates on the intermediate state and singlet states at sites $1$ and $5$, respectively, transforming them into triplet states, the pair hopping of dimers occurs. 
Defining the second-order pair-hopping amplitudes of the upper and lower processes shown in Fig.~\ref{fig:7} as $t_{\rm{s-s}}$ and $t_{\rm{t-t}}$, respectively, we can write 
\begin{align}
t=&t_{\rm{s-s}}+t_{\rm{t-t}}
\label{eq:14}.
\end{align}
The details of the calculation of the second-order perturbation energies (Eqs.~(\ref{eq:11})--(\ref{eq:13})) and pair-hopping amplitude (Eq.~(\ref{eq:14})) are provided in the Appendix.

\section{IV.~Results and discussion}

\begin{figure}[b]
\begin{center}
\includegraphics[width=10.0cm,clip]{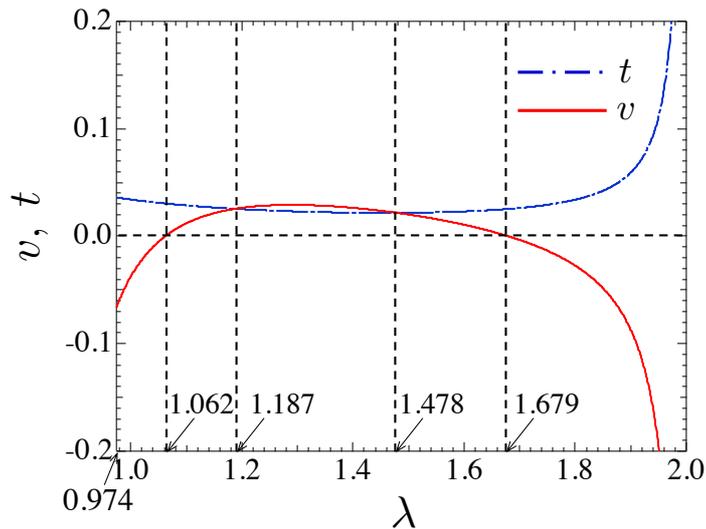}
\caption{(Color online) Calculation results for the dependence of $v$ and $t$ on $\lambda$, where the unit of $v$ and $t$ is $\Delta^2/J$.}
\label{fig:8}
\end{center}
\end{figure}

\begin{figure}[h]
\begin{center}
\includegraphics[width=10.0cm,clip]{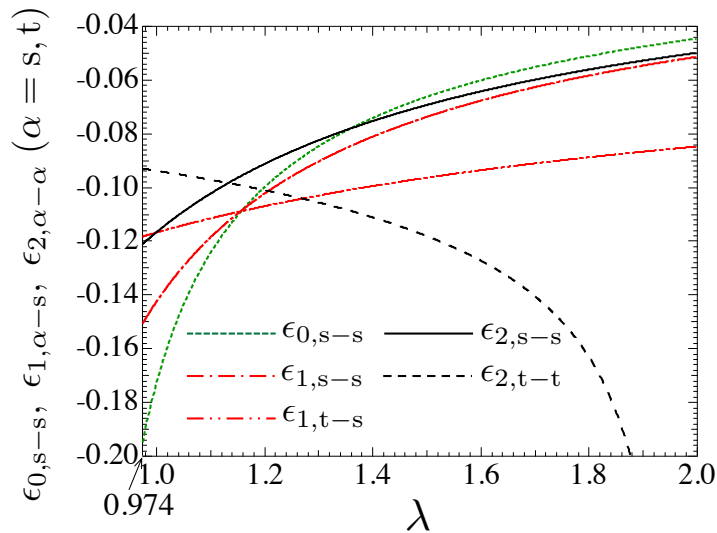}
\caption{(Color online) Calculation results for dependence of $\epsilon_{0,\rm{s-s}}$, $\epsilon_{1,\rm{s-s}}$, $\epsilon_{1,\rm{t-s}}$, $\epsilon_{2,\rm{s-s}}$, and $\epsilon_{2,\rm{t-t}}$ on $\lambda$, where the unit of these energies is $\Delta^2/J$.}
\label{fig:9}
\end{center}
\end{figure}

\begin{figure}[h]
\begin{center}
\includegraphics[width=10.0cm,clip]{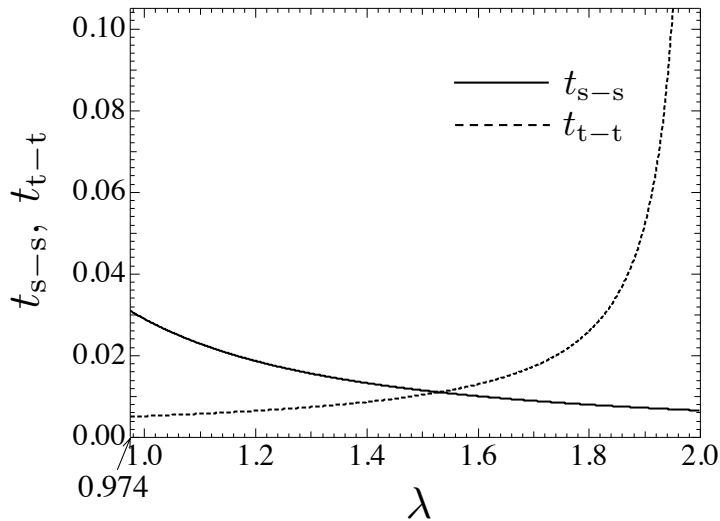}
\caption{Calculation results for dependence of $t_{\rm{s-s}}$ and $t_{\rm{t-t}}$ on $\lambda$, where the unit of $t_{\rm{s-s}}$ and $t_{\rm{t-t}}$ is $\Delta^2/J$.}
\label{fig:9-2}
\end{center}
\end{figure}

The numerical calculation results for the dependence of $v$ and $t$ on $\lambda$ are shown in Fig.~\ref{fig:8}.
The horizontal axes are in the range of $0.974<\lambda<2.0$, where the MDTD states are stabilized and the square-lattice dimer-covering states are constructed. 
First, it should be noted that at $\lambda=1.187$ and 1.478, we obtain the RK point $(v=t>0)$, i.e., a stabilized RVB state.   
Furthermore, Fig.~\ref{fig:8} shows that the attractive (repulsive) interaction, $v<0$ ($v>0$), is obtained in the ranges of $0.974<\lambda<1.062$ and $1.679<\lambda<2.0$ ($1.062<\lambda<1.679$). 
To investigate these behaviors in detail, Fig.~\ref{fig:9} shows the numerical calculation results for the dependence of $\epsilon_{0,\rm{s-s}}$, $\epsilon_{1,\rm{s-s}}$, $\epsilon_{1,\rm{t-s}}$, $\epsilon_{2,\rm{s-s}}$, and $\epsilon_{2,\rm{t-t}}$ on $\lambda$ and the dependence of $t_{\rm{s-s}}$ and $t_{\rm{t-t}}$ on $\lambda$ is shown in Fig.~\ref{fig:9-2}. 
Figure~\ref{fig:9} shows that the energy gain of a plaquette with no dimer, $\epsilon_{0,\rm{s-s}}$, is increasing as $\lambda$ approaches 0.974. 
Therefore, as shown in Fig.~\ref{fig:8}, the attractive interaction becomes larger as $\lambda$ approaches 0.974. 
On the other hand, the energy gain of a plaquette with two dimers, $\epsilon_{2,\rm{t-t}}$, diverges to $-\infty$ at $\lambda=2.0$, which results in the divergence of $v\to-\infty$ at $\lambda=2.0$, as shown in Fig.~\ref{fig:8}. 
The pair-hopping amplitude, $t$, diverges to $+\infty$ at $\lambda=2.0$ because $t_{\rm{t-t}}$ diverges to $+\infty$ at $\lambda=2.0$, as shown in Fig.~\ref{fig:9-2}.
The reason for the divergence at $\lambda=2.0$, which is a phase transition point in the original Hamiltonian in Eq.~(\ref{eq:2}), is that the energy denominators become zero (see Appendix).   
In the intermediate region of $\lambda$, the energy gain of a plaquette with one dimer is slightly larger than those of the others, resulting in a small repulsive interaction.  

\begin{figure}[t]
\begin{center}
\includegraphics[width=12.0cm,clip]{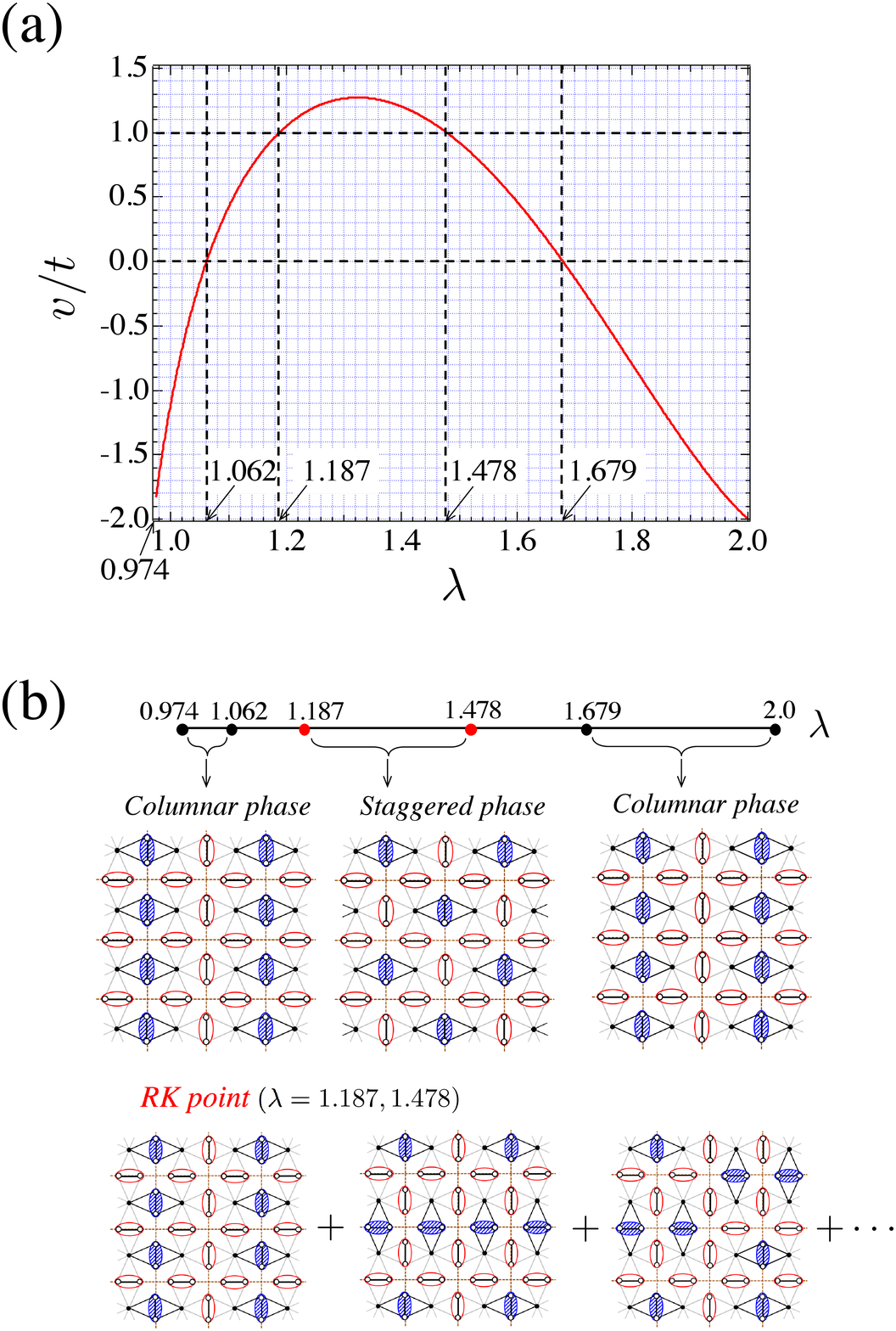}
\end{center}
\caption{(Color online) (a) Calculation results for dependence of $v/t$ on $\lambda$ and (b) phase diagram of $\lambda$.}
\label{fig:10}
\end{figure}

The numerical calculation results for the dependence of $v/t$ on $\lambda$ and the phase diagram of $\lambda$ are shown in Figs.~\ref{fig:10}(a) and (b), respectively.
At $\lambda=1.187$ and 1.478, $v/t=1$, which is the RK point that is obtained as detailed above. 
In the region of $1.187<\lambda<1.478$, $v/t>1$ is obtained, which suggests the stabilization of the staggered phase of the square-lattice QDM due to the generation of a large repulsive interaction between the dimers. 
On the other hand, in the ranges of $0.974<\lambda<1.062$ and $1.679<\lambda<2.0$, $v/t<0$ is obtained, which suggests the stabilization of the columnar phase of the square-lattice QDM due to the generation of an attractive interaction between dimers\cite{ref.book,ref.souzu}. 

\newpage
\section{V.~Summary}

We obtained a square-lattice QDM containing the RK point, as a second-order effective Hamiltonian, from spin-1/2 Heisenberg antiferromagnets on a DDSL by introducing further neighbor couplings in Eq.~(\ref{eq:5}). 
We found that by controlling the magnitude of the parameter $\lambda$ on the DDSL, it is possible to freely generate the states appearing in the square-lattice QDM. 
We believe that the most important result of the present study is the construction of a QDM containing the RK point from a model containing only quadratic and isotropic Heisenberg-type couplings. 
This is contrary to previous studies, where it has not been possible to construct a QDM from quantum spin systems without considering the model containing complicated multiple spin interactions such as those in Ref.~\onlinecite{ref.Fujimoto}. Therefore, we expect that our construction of QDMs from Heisenberg models provides a clear path toward experimental realization, such as a quantum simulator using an optical lattice\cite{ref.Bloch2008,ref.Bloch2017,ref.Bhuvanesh,ref.Fukuhara}. 
Furthermore, for triangular-lattice QDM, it has been reported that the RVB state emerges in a finite area of $0.8<v/t\le1$\cite{ref.triangular_qdm_2001,ref.triangular_qdm_2005}. 
Our method for constructing QDM can be adopted for arbitrary lattices, i.e., dimer-covering states of a corresponding lattice can be obtained if we replace each bond of an arbitrary lattice by the Hamiltonian in Eq.~(3) for $\lambda_{\rm{c}}<\lambda<2.0$\cite{ref.lambdac}. Therefore, if we consider a diamond-decorated triangular lattice\cite{ref.Hirose5} instead of a DDSL, the obtained effective Hamiltonian may provide the RVB state over a finite area of $\lambda$. 
We expect that the result in the present study is the first step toward such a concept.

\section{Acknowledgment}

The present author would like to thank Associate Professor Chitoshi Yasuda for careful reading of the manuscript. 
This work was supported by JSPS KAKENHI, Grant Number: JP17K05519. 

\section*{Appendix: \;Calculations of the second-order perturbation matrix elements}

\setcounter{equation}{0} 
\renewcommand{\theequation}{A.\arabic{equation}}
\setcounter{figure}{0} 
\renewcommand{\thefigure}{A.\arabic{figure}}
In this section, the details of the calculations of the second-order perturbation matrix elements are presented.  
We consider the upper process $\epsilon_{2,\rm{s-s}}$ shown in Fig.~6(a). When $V_{1,5}$ operates on the initial state $|s_{1};\phi^g_{2,4,3};s_{5};\phi^g_{0,6,7}\rangle$, we have 
\begin{align}
V_{1,5}|s_{1};\phi^g_{2,4,3};s_{5};\phi^g_{0,6,7}\rangle=-\frac{\Delta}{4}\left(|t^0\rangle_1|t^0\rangle_{5}-|t^+\rangle_1|t^-\rangle_{5}-|t^-\rangle_1|t^+\rangle_{5}\right)|\phi^g_{2,4,3};\phi^g_{0,6,7}\rangle
   \label{eq:A1}.
\end{align}
If the bond spin pairs at sites $1$ and $5$ are in triplet states, then a connected cluster $(0,1,2,3,4,5,6,7)$ appears in the intermediate state. 
For the unperturbed Hamiltonian, $h_{0,2}+h_{2,4}+h_{4,6}+h_{6,0}$, of this cluster, the eigenvalues and eigenstates can be written as $E_{0-7}^n+4J\lambda$ and $|\Psi^n\rangle_{0-7}$, respectively.
Therefore, the matrix elements can be written as 
\begin{align}
\langle\Psi^n_{0-7}|V_{1,5}|s_{1};\phi^g_{2,4,3};s_{5};\phi^g_{0,6,7}\rangle=&-\frac{\Delta}{4}\Large(\langle\Psi^n_{0-7}|t_1^0;t_5^0;\phi^g_{2,4,3};\phi^g_{0,6,7}\rangle-\langle\Psi^n_{0-7}|t_1^+;t_5^-;\phi^g_{2,4,3};\phi^g_{0,6,7}\rangle\notag \\
&-\langle\Psi^n_{0-7}|t_1^-;t_5^+;\phi^g_{2,4,3};\phi^g_{0,6,7}\rangle\Large)
 \label{eq:A2}.
\end{align}
The energy denominator of the intermediate state is given by
\begin{align}
(E_{0-7}^n+4J\lambda)-2J(\lambda-2)=E_{0-7}^n+J(2\lambda+4)
 \label{eq:A3}.
\end{align}
Thus, from Eqs.~(\ref{eq:A2}) and (\ref{eq:A3}), we obtain
\begin{align}
\epsilon_{2,\rm{s-s}}=&-\frac{\Delta^2}{16}\sum_{n}\Bigl(\langle\Psi^n_{0-7}|t_1^0;t_5^0;\phi^g_{2,4,3};\phi^g_{0,6,7}\rangle-\langle\Psi^n_{0-7}|t_1^+;t_5^-;\phi^g_{2,4,3};\phi^g_{0,6,7}\rangle\notag \\
&-\langle\Psi^n_{0-7}|t_1^-;t_5^+;\phi^g_{2,4,3};\phi^g_{0,6,7}\rangle\Bigr)^2\big/{\left\{E_{0-7}^n+J(2\lambda+4)\right\}}
\label{eq:A4}.
\end{align}

Similarly, $\epsilon_{2,\rm{t-t}}$, $\epsilon_{1,\rm{s-s}}$, $\epsilon_{1,\rm{t-s}}$, $\epsilon_{0,\rm{s-s}}$, $t_{\rm{s-s}}$, and $t_{\rm{t-t}}$ can be obtained by 
\begin{align}
\epsilon_{2,\rm{t-t}}=&\frac{\Delta^2}{J}\left\{\frac{1}{96(\lambda-2)}-\frac{1}{24}-\frac{1}{12(3-\lambda)}\right\}
 \label{eq:A5},
\end{align}
\begin{align} 
\epsilon_{1,\rm{s-s}}=&-\frac{\Delta^2}{16}\sum_{n}\bigl(\langle\Psi^n_{0-10}|t_3^0;t_7^0;\phi^g_{0,2,1};\phi^g_{4,6,5};\phi^g_{8,10,9}\rangle-\langle\Psi^n_{0-10}|t_3^+;t_7^-;\phi^g_{0,2,1};\phi^g_{4,6,5};\phi^g_{8,10,9}\rangle \notag \\
&-\langle\Psi^n_{0-10}|t_3^-;t_7^+;\phi^g_{0,2,1};\phi^g_{4,6,5};\phi^g_{8,10,9}\rangle\bigr)^2\big/\left\{E_{0-10}^n+J(2\lambda+6)\right\}
\label{eq:A6},
\end{align}
\begin{align} 
\epsilon_{1,\rm{t-s}}=&-\frac{\Delta^2}{48}\sum_{n}\Bigl({|\langle\Psi^n_{0,1,2,8,9,10,11}|t^+_{11};\phi^g_{0,2,1};\phi^g_{8,10,9}\rangle|}^2+{|\langle\Psi^n_{0,1,2,8,9,10,11}|t^-_{11};\phi^g_{0,2,1};\phi^g_{8,10,9}\rangle|}^2 \notag \\
&+{|\langle\Psi^n_{0,1,2,8,9,10,11}|t^0_{11};\phi^g_{0,2,1};\phi^g_{8,10,9}\rangle|}^2\bigr)\big/\left(E_{0,1,2,8,9,10,11}^n+6J\right)\notag \\
&-\frac{\Delta^2}{24}\sum_n\Bigl({|\langle\Psi^n_{0,1,2,8,9,10,11}|t^0_{11};\phi^g_{0,2,1};\phi^g_{8,10,9}\rangle|}^2+{|\langle\Psi^n_{0,1,2,8,9,10,11}|t^-_{11};\phi^g_{0,2,1};\phi^g_{8,10,9}\rangle|}^2\notag \\
&+{|\langle\Psi^n_{0,1,2,8,9,10,11}|t^+_{11};\phi^g_{0,2,1};\phi^g_{8,10,9}\rangle|}^2\Bigr)\big/\left\{E^n_{0,1,2,8,9,10,11}+J(\lambda+5)\right\}
\label{eq:A7},
\end{align}
\begin{align} 
\epsilon_{0,\rm{s-s}}=&-\frac{\Delta^2}{16}\sum_{n,n'}\Bigl({|\langle\Psi^n_{0-6}|t_3^0;\phi^g_{0,2,1};\phi^g_{4,6,5}\rangle\langle\Psi^{n'}_{8-14}|t_{11}^0;\phi^g_{8,10,9};\phi^g_{12,14,13}\rangle|}^2 \notag \\
&+{|\langle\Psi^n_{0-6}|t_3^+;\phi^g_{0,2,1};\phi^g_{4,6,5}\rangle\langle\Psi^{n'}_{8-14}|t_{11}^-;\phi^g_{8,10,9};\phi^g_{12,14,13}\rangle|}^2\notag \\
&+{|\langle\Psi^n_{0-6}|t_3^-;\phi^g_{0,2,1};\phi^g_{4,6,5}\rangle\langle\Psi^{n'}_{8-14}|t_{11}^+;\phi^g_{8,10,9};\phi^g_{12,14,13}\rangle|}^2\Bigr)\big/\left\{E_{0-6}^n+E_{8-14}^{n'}+J(2\lambda+8)\right\}
\label{eq:A8},
\end{align}
\begin{align} 
t_{\rm{s-s}}=&-\frac{\Delta^2}{16}\sum_{n}\bigl(\langle\Psi^n_{0-7}|t_3^0;t_7^0;\phi^g_{0,2,1};\phi^g_{4,6,5}\rangle-\langle\Psi^n_{0-7}|t_3^+;t_7^-;\phi^g_{0,2,1};\phi^g_{4,6,5}\rangle \notag \\
			 &-\langle\Psi^n_{0-7}|t_3^-;t_7^+;\phi^g_{0,2,1};\phi^g_{4,6,5}\rangle\bigr)
			 \times\bigl(\langle\Psi^n_{0-7}|t_1^0;t_5^0;\phi^g_{2,4,3};\phi^g_{0,6,7}\rangle-\langle\Psi^n_{0-7}|t_1^+;t_5^-;\phi^g_{2,4,3};\phi^g_{0,6,7}\rangle \notag \\
			 &-\langle\Psi^n_{0-7}|t_1^-;t_5^+;\phi^g_{2,4,3};\phi^g_{0,6,7}\rangle\bigr)/\left\{E_{0-7}^n+J(2\lambda+4)\right\}
\label{eq:A9},
\end{align}
and
\begin{align} 
t_{\rm{t-t}}=-\frac{\Delta^2}{192J(\lambda-2)}
\label{eq:A10}.
\end{align}

It can be seen, that the denominator of the first term of Eq.~(\ref{eq:A5}) and that of Eq.\;(\ref{eq:A10}) generate the divergence of $v\to-\infty$ and $t\to+\infty$\cite{ref.sign}, respectively, at $\lambda=2.0$, as shown in Fig.~8. 

The values of Eqs.~(\ref{eq:A4}), (\ref{eq:A6}), (\ref{eq:A7}), (\ref{eq:A8}), and~(\ref{eq:A9}) for each $\lambda$ are obtained by using the exact diagonalization method for the clusters consisting of 8, 11, 7, 14, and 8 sites, respectively. 
 
 \bibliography{your-bib-file}

\begin{thebibliography}{99}
\bibitem{ref.sign} In this study, the coefficient on the right side of Eqs.~(\ref{eq:A9}) and (\ref{eq:A10}) are determined to be negative, and as a result, both $t_{\rm{s-s}}$ and $t_{\rm{t-t}}$ have positive values, as shown in Fig.~\ref{fig:9-2}. However, it should be noted that the sign of the hopping parameter is not important because it can be absorbed into a phase factor of the dimer-covering states.
\bibitem{ref.lambdac} The phase boundary $\lambda_{\rm{c}}$ depends on the lattices. In the case of diamond-like decorated square (triangular) lattice, we obtain $\lambda_{\rm{c}}=0.974$ (0.988)\cite{ref.Hirose5}.
\end{thebibliography}

\end{document}